\documentclass[10pt]{article}
\usepackage{a4wide}                      
\usepackage{epsf}                        
\usepackage[dvips]{graphicx}
\usepackage{latexsym}                    
\usepackage{amsfonts}                    
\usepackage{amssymb}                     

\def\be{\begin{equation}}
\def\ee{\end{equation}}
\def\bea{\begin{eqnarray}}
\def\eea{\end{eqnarray}}

\def\vsp#1{\vspace{#1}}
\def\hsp#1{\hspace{#1}}
\def\vsp*#1{\vspace*{#1}}
\def\hsp*#1{\hspace*{#1}}

\def\part{\partial}

\def\makeatletter{\catcode`\@=11}
\makeatletter
\def\mathbox#1{\hbox{$\m@th#1$}}%
\def\math@ccstyles#1#2#3#4#5#6#7{{\leavevmode
      \setbox0\mathbox{#6#7}%
      \setbox2\mathbox{#4#5}%
      \dimen@ #3%
      \baselineskip\z@\lineskiplimit#1\lineskip\z@
      \vbox{\ialign{##\crcr
             \hfil \kern #2\box2 \hfil\crcr
             \noalign{\kern\dimen@}%
             \hfil\box0\hfil\crcr}}}}
\def\mathaccstyles{\math@ccstyles\maxdimen}
\def\maththroughstyles{\math@ccstyles{-\maxdimen}}
\def\unity%
 {\maththroughstyles{.45\ht0}\z@\displaystyle {\mathchar"006C}\displaystyle 1}
\begin{document}

\rightline{UG-FT-214/07}
\rightline{CAFPE-84/07}
\rightline{FFUOV-07/02}
\rightline{hep-th/0701151}
\rightline{January 2007}
\vspace{1truecm}

\begin{center}
{\huge \bf Adding magnetic flux to } \\ [.4cm]
{\huge \bf the baryon vertex}
\end{center}

\vspace{1truecm}

\centerline{
    {\large \bf Bert Janssen${}^{a,}$}\footnote{E-mail address: 
                                  {\tt bjanssen@ugr.es} },
    {\large \bf Yolanda Lozano${}^{b,}$}\footnote{E-mail address:
                                  {\tt yolanda@string1.ciencias.uniovi.es}}
    {\bf and} 
    {\large \bf Diego Rodr\'{\i}guez-G\'omez${}^{c,}$}\footnote{E-mail address:
                                   {\tt drodrigu@princeton.edu}} 
                                                            }
\vspace{.4cm}
\centerline{{\it  ${}^a$Departamento de F\'{\i}sica Te\'orica y del Cosmos and}}
\centerline{{\it Centro Andaluz de F\'{\i}sica de Part\'{\i}culas Elementales}}
\centerline{{\it Universidad de Granada, 18071 Granada, Spain}} 

\vspace{.4cm}
\centerline{{\it ${}^b$Departamento de F{\'\i}sica,  Universidad de Oviedo,}}
\centerline{{\it Avda.~Calvo Sotelo 18, 33007 Oviedo, Spain}}

\vspace{.4cm}
\centerline{{\it ${}^c$ Department of Physics, Princeton University,}}
\centerline{{\it Princeton, NJ 08540, U.S.A.}}

\vspace{1truecm}

\centerline{\bf ABSTRACT}
\vspace{.5truecm}

\noindent
We generalise the baryon vertex configuration of AdS/CFT by adding 
magnetic field on its worldvolume, dissolving D-string charge. A careful analysis of the 
configuration shows that there is an upper bound on the number of dissolved strings. We provide 
a microscopical description of this configuration in terms of a dielectric effect
for the dissolved strings.\footnote{Talk given by B.J. at the RTN Network Meeting in Napoli, in 
     Octobre 2006.}

\newpage

\section{Introduction}

The AdS/CFT correspondence conjectures a remarkable duality between strings moving in an 
$AdS_5 \times S^5$ background with $N$ units of 4-form R-R flux and $D=4, \ {\cal N} = 4$ 
SYM in flat space with gauge group $SU(N)$ \cite{Maldacena}. In particular, the conjecture 
states that quark and anti-quarks in the $SU(N)$ gauge theory manifest themselves in 
$AdS_5 \times S^5$  as the endpoints of open strings ending on the boundary of the $AdS_5$. 

It is clear then that one can form a $q\bar{q}$ state with a single U-shaped string coming 
from the boundary, entering into the bulk, with the apex at a distance $u_0$ from the centre 
of the $AdS_5$ and then going back to the boundary. This configuration is seen  on the SYM 
side as a $q\bar{q}$ pair (a meson), whose energy is computed by means of a Wilson loop, 
while on the gravity side the energy is computed by minimising the worldsheet area of the 
string ending on the loop \cite{Maldacena2, RY}.

One would expect that, besides the meson, there should be a mechanism to form a baryon 
configuration, {\it i.e.} an anti-symmetric, colourless, finite-energy bound state of $N$ 
non-dynamical quarks. In \cite{Witten} the gravitational dual of such bound state of quarks 
was found in terms of a D5-brane wrapping the $S^5$ part of the $AdS_5 \times S^5$ background.
On this D5-brane there are $N$ F-strings attached, stretching from the D5-brane to the 
boundary of $AdS_5$, whose endpoints are regarded on the SYM side as a bound state of $N$ 
quarks. It can be shown \cite{Witten} that the associated wave function satisfies the 
required symmetry properties. 

The best way to see that the above configuration is indeed the baryon vertex is by looking at the 
Chern-Simons term of the D5-brane, wrapped around the $S^5$.  Since in $AdS_5 \times S^5$ there 
is no 6-form R-R potential to which the probe brane can couple, the only non-zero term in the 
Chern-Simons action is  the 4-form R-R field $C^{(4)}$ coupled to the BI field strength $F=dA$. 
In our specific setting, the only non-zero contribution is that of the coupling of the magnetic 
part of $C^{(4)}$ to the electric component of $F$ . Integrating by parts, this term can 
then be rewritten as
\begin{equation}
S_{CS}\ = \ -T_5\int_{{\mathbb{R}}\times S^5} P[C^{(4)}]\wedge F
      \ = \ T_5\int_{{\mathbb{R}}\times S^5} P[G^{(5)}]\wedge A,
\label{C^4F}
\end{equation}
where $G^{(5)} = dC^{(4)}$ is the R-R 5-form field strength. In $AdS_5 \times S^5$  we have that  
$\int_{S^5}G^{(5)}=4\pi^2 N$ (in units where $2\pi l_s^2=1$), such that with the Ansatz 
$A = A_t (t) dt$, it is clear that the coupling (\ref{C^4F}) factorises as
\begin{equation}
\label{SCSD5}
S_{CS}=T_5\int_{S^5}G^{(5)}\int dt A_t\ =\ N T_1 \int dt A_t,
\end{equation}
where we have taken into account that the tension of the D5-brane and the tension of a string 
are related by $4\pi^2 T_5=T_1$. Therefore, the coupling (\ref{C^4F}) is 
inducing $N$ units of electric BI charge on the D5-brane. However we have to check whether the 
Ansatz for $A$ is consistent with the equations of motion of the total D5-brane system. 
The equation of motion of $A$ is given by $N T_1 = 0$, such that the above construction is only 
consistent if the total electric BI charge on the D5-brane is zero. Yet, there is a way of 
inducing a non-zero BI electric 
charge in the worldvolume of the D5-brane, by cancelling this charge with the charge induced 
by the endpoints of $N$ open fundamental strings (with appropriate orientation) stretching 
between the D5-brane and the boundary of the $AdS$ space. The full dynamics of the configuration 
is therefore not only described by the $D5$-brane action, but also by the action for the 
open strings, consisting of $N$ copies of the Nambu-Goto action $S_{F1}$ and a boundary term 
contribution $T_1\int A_t dt$ from the endpoints. Note that the contribution from the open string 
endpoints cancels exactly the Chern-Simons term in the D5-brane action, such that the total
system is described by \cite{BISY} 
\begin{equation}
S_{\rm total} \ = \ S_{DBI}   +  N T_1 \int dt A_t \ \ + \ \ N S_{F1}  -  N T_1\int dt A_t
              \ = \ S_{DBI}\ \ + \ \ N S_{F1}.
\label{S=S5+S1}
\end{equation}
The configuration that we have just described is the so-called baryon vertex. Since the $N$ 
F-strings, stretching from the D5-brane all the way to the $AdS$ boundary, have the same 
orientation, the dual configuration on the CFT side corresponds to the bound state of $N$ 
(anti)quarks, which is gauge invariant and antisymmetric under the interchange of any two 
quarks \cite{Witten}.

We will show that the baryon vertex can be generalised by adding a new quantum number, representing 
magnetic flux. The key point is to realise that $S^5$ can be seen as an $S^1$ bundle over $CP^2$. 
The $S^1$ fibre is a non-trivial $U(1)$ gauge bundle on the $CP^2$ base, and this allows to switch 
on a magnetic BI field  on the worldvolume of the D5-brane, proportional to the curvature tensor 
of the fibre connection. We will show that the magnetic flux leads to surprising dynamics and a 
complementary description of the baryon vertex as a dielectric effect. The results mentioned in 
this letter are discussed in detail in \cite{JLR4}.

\section{A generalisation of the baryon vertex}
It is well known that $S^5$ can be regarded as a $U(1)$ fibre over $CP^2$ with a non-trivial fibre 
connection $B$. In these $S^5$ fibre coordinates the $AdS_5\times S^5$ background then reads
\begin{eqnarray}
&& ds^2\ =\ \frac{u^2}{L^2}\eta_{ab} dx^adx^b\  + \ \frac{L^2}{u^2}du^2 \ 
      + \ L^2 \Bigl( (d\chi-B)^2+ds_{CP^2}^2\Bigr),\nonumber \\
&&C_{abcd} = L^{-4}u^4 \epsilon_{abcd}\ , \hspace{2cm}
C_{\varphi_2\varphi_3 \varphi_4\chi} = \frac{1}{8} L^4 \sin^4 \varphi_1 \sin \varphi_2,
\label{background}
\end{eqnarray}
where $ds_{CP^2}^2$ stands for the Fubini-Study metric on $CP^2$ (with coordinates $\varphi_i$) and 
$\chi$ is taken along the $U(1)$ fibre. The fibre connection $B$ satisfies the following properties 
\cite{Trautman}
\begin{equation}
dB={}^\star(dB), \hspace{2cm}
\int_{CP^2} dB \wedge dB =4\pi^2\, , 
\label{propertiesB}
\end{equation}
where the Hodge star is taken with respect to the Fubini-Study metric on $CP^2$. In particular,
one sees that $B$ induces a non-trivial instanton number in $CP^2$  \cite{Trautman, HP}.
In these coordinates the baryon vertex consists of the D5-brane wrapped around the $S^5$ and the 
fundamental strings laying in the $u$-direction of $AdS_5$ \cite{RY}. Besides the electric components 
of the BI field strength of the previous section, one could think of turning on also magnetic 
components. Due to (\ref{propertiesB}), it is natural to take the magnetic components living in the 
$CP^2$ and proportional to the curvature tensor of the $U(1)$ fibre connection, $F = 2n \, dB$.
With this Ansatz $F$ satisfies the same properties (\ref{propertiesB}) as the fibre connection,  
namely it is selfdual and
\begin{equation}
\int_{CP^2}F\wedge F=  8\pi^2 n^2\, .
\label{FF=n}
\end{equation}
With this
choice for the BI field strength it is clear that there are no other couplings
in the Chern-Simons action besides the ones we already considered in (\ref{C^4F}). The 
Born-Infeld action however is given by
\begin{eqnarray}
S_{DBI}\ = \ -T_5\int d^6 \xi \ \frac{u}{L}\ 
                      \sqrt{\det \Bigl( g_{\alpha\beta} + F_{\alpha\beta}\Bigr)} 
       \ =\  -T_5  \int d^6 \xi \  u\ \sqrt{g_{S^5}} \ 
             \Bigl( L^4 + 2 F_{\alpha\beta}F^{\alpha\beta}\Bigr),
\label{DBI}
\end{eqnarray}
where we have used that $F$ is selfdual, such that the determinant under the square root is a perfect 
square. The Ansatz for $F$ is consistent with the action (\ref{DBI}), as is reflected in the fact that 
the equations of motion for the magnetic components of $F$ are given by $dF= 0$. Finally, substituting 
the expression for $F$ in the action and integrating over the $S^5$ directions we obtain the following 
expression for the energy of the spherical D5-brane:  
\begin{equation}
E_{D5}\ = \ 8\pi^3  T_5\ u \ \Big(n^2 +\frac{L^4}{8}\Big).
\label{ED5}
\end{equation}
Note that this energy consists of two parts: one contribution from the tension of the 
5-brane wrapped around the $S^5$, proportional to $L^4$ and one from the magnetic flux of the
BI vector, proportional to $n^2$. 

The magnetic components of $F$ induce a non-zero instanton number $n^2$ on the worldvolume of the
D5-brane, since integrating the Chern-Simons coupling over the $CP^2$ directions, one obtains
\begin{equation}
S_{CS}\ =\ \frac{1}{2}T_5\int_{\ensuremath{\mathbb{R}}\times S^5} P[C^{(2)}]\wedge F\wedge F 
      \ = \ n^2 T_1\int_{\ensuremath{\mathbb{R}} \times S^1} P[C^{(2)}]\ ,
\label{FFcoupling2}
\end{equation}
where we have used again that $T_1 = 4\pi^2 T_5$. Even though in $AdS_ 5 \times S^5$ $C^{(2)}$ is 
zero, this coupling indicates that the magnetic flux is  inducing $n^2$ D-string charge in the 
configuration. These strings are wound around the fibre direction $\chi$ and dissolved in the 
$CP^2$. Note that the total energy of the configuration (\ref{ED5}) is the sum of the energy of 
the D5 and the D1-branes, which hints at that we are dealing with a threshold BPS bound state.

\section{A bound on the magnetic flux}

It was argued in \cite{BISY} that in order to analyse the stability of the baryon vertex
in the $u$-direction, one has to consider the influence of the external F-strings. 
The energy $E$ of the baryon vertex is inversely proportional to the distance $\ell$ between 
the quarks, with a negative proportionality constant, such that the baryon vertex is indeed 
stable under perturbations in $u$. In this subsection we will perform the same calculation in 
\cite{BISY}, but taking into account the effect of the non-zero magnetic flux on the D5-brane. 

The equations of motion of F-strings in the baryon vertex are twofold: the bulk equation 
of motion for the strings and the boundary equation of motion, which contains a term 
coming from the D5-brane. Parametrising the worldsheet of the F-strings by
 $\{t,x\}$ and the position in $AdS$ by $u=u(x)$, we obtain
\begin{eqnarray}
\frac{u^4}{\sqrt{(u')^2+\frac{u^4}{L^4}}}\ = \ {\rm const}, 
\hspace{2cm}
\frac{u'_0}{\sqrt{(u'_0)^2+\frac{u_0^4}{L^4}}}
           \ = \ \frac{\pi L^4}{4N}\Bigl(1+\frac{8n^2}{L^4}\Bigr),
\label{boundary}
\end{eqnarray}
for the bulk and the boundary respectively, with $u_0$ the position of the baryon vertex in the 
holographic direction and  $u'_0 = u'(u_0)$. The equations (\ref{boundary}) can be combined 
into a single one,
\begin{equation}
\frac{u^4}{\sqrt{(u')^2+\frac{u^4}{L^4}}}\ =\ \beta \ u_0^2L^2, 
\hspace{2cm}
{\rm with} \ \ \beta^2\ = \ 1 - \frac{1}{16}\Bigl( 1 +\frac{8\pi n^2}{N}\Bigr)^2.
\end{equation}
In the absence of magnetic BI flux on the worldvolume, $\beta =\sqrt{15/16}$, as in \cite{BISY}. 
However, in general for non-zero $n^2$, we have to make sure that $\beta$ is real (as $u$ is real), 
which implies that $n^2 \le {{3N}/{8\pi}}$. Surprisingly, we find that there is a bound on the 
number of D-strings that can be dissolved in the configuration, which depends on the number of 
D3-branes that source the background. 

\begin{figure}
\includegraphics[width=65mm]{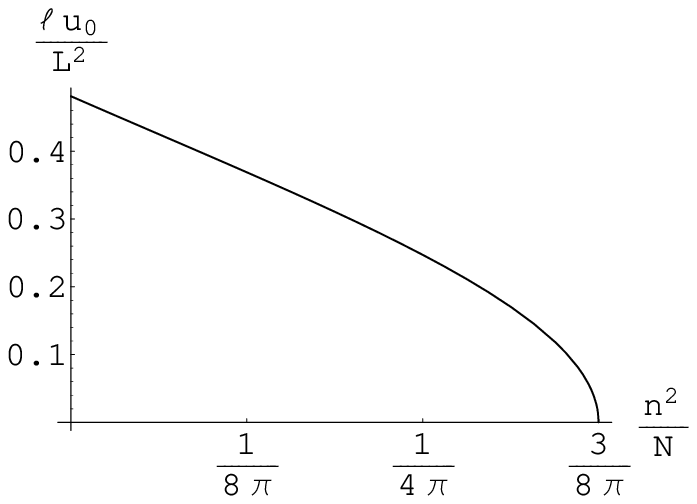}
\hfil
\includegraphics[width=65mm]{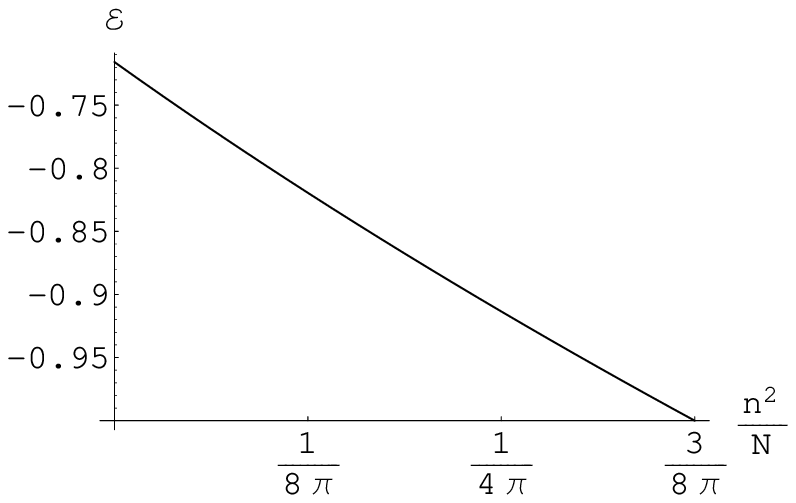}
\caption{Radius $\ell$ (in units of $L^2/u_0$) and the energy $E$ (in units of $u_0$) of the 
baryon vertex as a function of ${n^2}/{N}$.}
\label{plots}
\end{figure}

Integrating the equation of motion, we find that the size $\ell$ and the energy $E$ of the
baryon are given by
\begin{equation}
\ell=\frac{L^2}{u_0}\int_1^{\infty} dy\frac{\beta}{y^2\sqrt{y^4-\beta^2}}, 
\hspace{2cm}
E=T_1 u_0 \Bigr\{
    \int_1^{\infty}dy \Bigl[\frac{y^2}{\sqrt{y^4-\beta^2}}-1 \Bigr]  - 1 \ \Bigr\},
\end{equation}
with $y= u/u_0$. These integrals can be solved in terms of hypergeometric functions \cite{BISY}. 
In Figure \ref{plots} we have plotted the radius and the energy of the baryon as a function of 
${n^2}/{N}$. The plots reveal that the size of the baryon vertex goes to zero as we approach the 
bound on $n^2$, making it impossible to continue beyond the bound. The expression for the energy 
has the same form as the expression in \cite{BISY} and indeed takes the same value for $n=0$. 
In particular, the dependence on $\sqrt{g^2N}$ and on $u_0$ is unaltered, as expected by conformal 
invariance. Notice that also here the energy of the configuration is only well defined for ${n^2}$ 
inside the allowed interval. 

The fact that we find a bound on the number of dissolved D1-branes due to the dynamics of the 
F-strings, is quite surprising. The bound is probably related to the stringy exclusion principle 
of \cite{MaldaStrom}. Our configuration carries a non-zero winding number in the fibre direction 
of the $S^5$, which in terms of the dual field theory will manifest itself as a specific charge  
of the $SU(3)$ R-symmetry group. As these charges are bounded due to conformal invariance, one 
expects to find a bound on the magnetic flux. 

\section{A microscopical description in terms of D1-branes}

The fact that the magnetic flux induces D1-brane charge through the coupling (\ref{FFcoupling2}), 
suggests a close analogy with the dielectric effect described in \cite{Emparan, Myers}. In this 
section we will show that it is indeed possible to give an alternative, microscopic description 
of the baryon vertex, in terms of a fuzzy spherical D5-brane built up out of dielectrically 
expanded D1-branes.

The action describing the dynamics of $n^2$ coinciding D1-branes is the
non-Abelian action given in \cite{Myers}, which for the $AdS_5 \times S^5$
background reduces to the form
\begin{eqnarray}
\label{nD1}
S_{n^2D1} &=& - T_1 \int d^2\xi \ {\rm STr} \Bigl\{
       \sqrt{\Bigl|{\rm det}\Bigl(P[g_{\mu\nu} + g_{\mu i}(Q^{-1} -\delta)^i{}_j 
                             g^{jk}g_{k\nu}]\Bigr) 
                                 \det Q\Bigr| \ }\Bigr\}
\end{eqnarray}
where $g_{\mu\nu}$ is the metric in  $AdS_5 \times S^5$ and 
$Q^i{}_j = \delta^i_j + i [X^i, X^k] g_{kj}$. Inspired by the coupling (\ref{FFcoupling2}) in 
the D5-brane calculation we wind the D-strings around the $U(1)$ fibre direction $\chi$ and 
let them expand into the $CP^2$. In this way we obtain a fuzzy version of the $S^5$ as an 
Abelian $U(1)$ fibre over a fuzzy $CP^ 2$ \cite{JLR3}. 

A  fuzzy version of $CP^2$ (see for example \cite{ABIY}) can  be obtained by  the 
set of matrix coordinates 
\begin{equation}
X^i=\frac{T^i}{\sqrt{(2n^2-2)/3}},  \hspace{2cm}
[X^i, X^j] = \frac{i f^{ijk}}{\sqrt{(2n^2-2)/3}} X^k,
\label{X(T)}
\end{equation}
with $T^i$ the generators of $SU(3)$ in the $n^2$-dimensional
irreducible representations $(k,0)$ or $(0,k)$, with 
$n^2=(k+1)(k+2)/2$ (see \cite{JLR3,ABIY} for more details) and 
with $f^{ijk}$ the structure constant of $SU(3)$ in the algebra of
the Gell-Mann matrices $[\lambda^i,\lambda^j]=2i f^{ijk} \lambda^k$.

Substituting the non-commutative Ansatz above in the action (\ref{nD1}) we find
\begin{eqnarray}
S_{n^2D1} \ = \ - T_1 \int dt d\chi \ u \ 
               {\rm STr}\Bigl\{ \unity +\frac{L^4}{4(2n^2-2)}\unity\Bigr\}  
       \ = \ -2\pi n^2  T_1  \ \int dt \ u 
                           \ \Bigl( 1 +  \frac{L^4}{8(n^2-1)}\Bigr)\, ,
\end{eqnarray}
since remarkably ${\rm det}Q=(\unity + \frac{L^4}{4(2n^2-2)} \unity)^2$ in the large $n^2$ 
limit. It is to be emphasised that the fact that the $\det Q$ is a perfect square is the 
microscopical analogous of the perfect square that we obtained for the DBI action in the 
macroscopic case. The energy of the $n^2$ expanded D1-branes is then given by
\begin{equation}
E_{n^2D1} =  2\pi u  T_1 \ \Bigl( n^2 + \frac{n^2L^4}{8(n^2-1)}\Bigr). 
\label{EnD1}
\end{equation}
Taking into account that the tensions of the D1- and the D5-brane are related by
$T_1 = 4\pi^2 T_5$, it is easy to see that in the limit where the number of D1-branes 
$n^2\rightarrow \infty$, the above expression reduces to the energy
of the macroscopic D5-brane, given by (\ref{ED5}).

It is also possible to see the external F-strings in the microscopical description, by looking 
at the CS action for coincident D-strings. The coupling with the R-R 4-form and the $U(n^2)$ BI 
field strength ${\cal F} = d{\cal A} + [{\cal A}, {\cal A}]$ is, after partial integration 
and in the gauge ${\cal A}_\chi = 0$, given by
\begin{equation}
S_{CS} =-\frac{T_1}{4}\int dt d\chi \ {\rm STr}\Bigl\{[X^i,X^j][X^k,X^l]
                        G^{(5)}_{\chi ijkl}{\cal A}_t\Bigr\}
       = \frac{L^4 T_ 1}{2(n^2-1)} \int dtd\chi\ {\rm STr} \Bigl\{{\cal A}_t \Bigl\},
\label{CSD1}
\end{equation}
where we have used that  in the non-commutative coordinates introduced in (\ref{X(T)}), $G^{(5)}$ 
is given by \cite{JLR3} $G^{(5)}_{\chi ijkl}=L^4 f_{[ij}^m f_{kl]}^n X^m X^n$. In analogy with 
the Abelian case, we can take as an Ansatz for the BI vector, ${\cal A}=A_t (t) \unity dt$, such
that after integrating over $\chi$, we find finally that
\begin{equation}
S_{CS} =  \frac{n^2}{n^2-1}\ N T_1 \int dt \  A_t.   
\label{SCSD1FULL}
\end{equation}
The coupling (\ref{CSD1}) is therefore inducing, in the large $n^2$  limit, $N$ BI charges in the 
configuration. These charges have to be cancelled by $N$ fundamental strings ending on the D1-brane 
system. The dielectric coupling to $C^{(4)}$ in (\ref{CSD1}) will then take care that these 
strings are expanded over the full $S^5$.



\end{document}